\documentclass[12pt,nohyper,letterpaper]{JHEP3}         
\usepackage{amssymb,amsfonts,amsmath}

\usepackage{epsfig}

\def\be{\begin{eqnarray}}
\def\ee{\end{eqnarray}}
\newcommand{\nn}{\nonumber}
\newcommand\para{\paragraph{}}
\newcommand{\ft}[2]{{\textstyle\frac{#1}{#2}}}
\newcommand{\eqn}[1]{(\ref{#1})}

\def\Dslash{\,\,{\raise.15ex\hbox{/}\mkern-12mu D}}
\def\Dbarslash{\,\,{\raise.15ex\hbox{/}\mkern-12mu {\bar D}}}
\def\delslash{\,\,{\raise.15ex\hbox{/}\mkern-9mu \partial}}
\def\delbarslash{\,\,{\raise.15ex\hbox{/}\mkern-9mu {\bar\partial}}}
\def\pslash{\,\,{\raise.15ex\hbox{/}\mkern-9mu p}}
\def\calDslash{\,\,{\raise.15ex\hbox{/}\mkern-12mu {\cal D}}}

\def\lae{\mathrel{\mathop{\smash{\lower .5 ex \hbox{$\stackrel<\sim$}}}}}
\def\lae{\mathrel{\mathop{\smash{\lower .5 ex \hbox{$\stackrel>\sim$}}}}}

\def\Dslash{\,\,{\raise.15ex\hbox{/}\mkern-13mu D}}
\def\Dbarslash{\,\,{\raise.15ex\hbox{/}\mkern-12mu {\bar D}}}
\def\delslash{\,\,{\raise.15ex\hbox{/}\mkern-10mu \partial}}
\def\delbarslash{\,\,{\raise.15ex\hbox{/}\mkern-9mu {\bar\partial}}}
\def\pslash{\,\,{\raise.15ex\hbox{/}\mkern-11mu p}}
\def\qslash{\,\,{\raise.15ex\hbox{/}\mkern-9mu q}}
     \def\kslash{\,\,{\raise.15ex\hbox{/}\mkern-11mu k}}
\def\eslash{\,\,{\raise.15ex\hbox{/}\mkern-9mu \epsilon}}
\def\calDslash{\,\,{\rais.15ex\hbox{/}\mkern-12mu {\cal D}}}

\title{Berry Phase and Supersymmetry}
\author{Julian Sonner\footnote{On leave from DAMTP and Trinity College, Cambridge, {\tt j.sonner@damtp.cam.ac.uk}}\\
Center for Theoretical Physics, \\
Massachusetts Institute of Technology\\
Cambridge, MA 02139, USA\\
Email: {\tt sonner@mit.edu}
}
\author{David Tong\\
Department of Applied Mathematics and Theoretical Physics, \\
University of Cambridge, UK\\
Email: {\tt d.tong@damtp.cam.ac.uk}
}

\preprint{MIT-CTP 3989}
\abstract{We study the constraints of supersymmetry on the non-Abelian holonomy given by $U=P \exp(i\int A)$, the path-ordered exponential of a connection $A$. For theories with four supercharges, we show that $A$ satisfies the tt* equations if it is a function of chiral multiplets. In contrast, when $A$ is a function of vector multiplets, it satisfies the
Bogomolnyi monopole equations. We describe applications of these results to the Berry connection in supersymmetric quantum mechanics.}

\begin{document}
\pagestyle{plain} \setcounter{page}{1}
\newcounter{bean}
\baselineskip16pt

\subsection*{Introduction}

Often in physics we are interested in the holonomy of a state as we move along a path $\Gamma$ in some space ${\cal M}$. Such a holonomy is typically governed by the path-ordered exponential of a non-Abelian connection $A$ over ${\cal M}$,
\be
U = P\exp\left(i\int_\Gamma A \right) \label{holo}
\ee
In this short note we study the restrictions on the connection $A$ due to  supersymmetry. Specifically, we focus on situations where the coordinates $x^i$ over ${\cal M}$ can be thought of as the bosonic components of a  supermultiplet. In this case, the connection $A=A_i(x)\dot{x}^i\,dt$ is merely the leading order term in a Lagrangian,
\be L = A_i(x)\,\dot{x}^i + \ldots \label{lag}\ee
where $\ldots$ denote terms involving fermions and auxiliary fields which form the supersymmetric completion of the connection.  There is something unfamiliar about Lagrangians of this type: they are matrix-valued functions of scalar fields. This is to be contrasted with the more familiar quantum mechanical matrix models, where the Lagrangian is a scalar function of matrix-valued fields.

\para
Below, we study the conditions for the matrix-valued Lagrangian $L$ to be invariant under ${\cal N}=(2,2)$ supersymmetry (that is, the dimensional reduction of ${\cal N}=1$ supersymmetry in 4d). We restrict our attention to scalar fields that live in chiral multiplets or vector multiplets. When $A$ is a function of complex, chiral multiplet scalars, we show that supersymmetry restricts the connection to satisfy the tt* equations of \cite{C1,tt*}. In contrast, when $A$ depends on the triplet of scalars that live in a vector multiplet, the connection $A$ is constrained to obey the Bogomolnyi monopole equation \cite{bog}.

\para
At the end of this paper, we present an application of these results to computing the non-Abelian Berry phase in supersymmetric quantum mechanics. This was the original context in which the tt* equations were first discovered \cite{C1} and the method of this paper gives a particularly simple derivation. More recently, we have studied examples of quantum mechanics in in which the Berry connection obeys the Bogomolnyi monopole equations \cite{us1,us2}. It was conjectured in \cite{us2} that the Bogomolnyi equations are, more generally, analogous to the tt* equations for vector multiplet parameters. The results of this paper prove this conjecture.

\para
Recent related work has examined the Berry phases that arise in D-branes and supersymmetric black holes \cite{us44,d0,bh1,bh2}. We expect the results of this paper to be relevant to this study. The methods here should also be applicable to systems exhibiting different amounts of supersymmetry.

\subsubsection*{The Invariance of a Matrix}

The first question that we have to answer is: what does it mean for a matrix-valued Lagrangian $L$ to be invariant under a symmetry? In the familiar situation, where the Lagrangian is a scalar-valued function, a symmetry is any transformation under which the Lagrangian changes by a total derivative,
\be \delta L = \frac{d\Theta}{dt} \label{td}\ee
for some function $\Theta$. However, as explained in \cite{kentaro}, this is no longer the appropriate condition when $L$ is matrix-valued. The object of interest is now the time-ordered exponential,
\be U(t_i,t_f) = T\exp\left(i\int_{t_i}^{t_f} L(t)\, dt\right)\label{done}\ee
For concreteness $L$ is assumed to be a Hermitian $N\times N$ matrix, but more generally can be valued in any Lie algebra. Varying the Lagrangian results in a variation of the holonomy,
\be \delta U(t_f,t_i) = i \int_{t_i}^{t_f} U(t_i,t)\,\delta L (t)\,U(t,t_f) \, dt \label{vu}\ee
If the Lagrangian changes by a total derivative, as in \eqn{td}, the change in the holonomy has no particularly special properties. Instead, a transformation is said to be a symmetry if the integrand in \eqn{vu} is a total derivative: $\ft{d}{dt}[U(t_i,t)\Theta(t)U(t,t_f)]$. This holds if $\delta L$ is a total covariant derivative,
\be \delta L  = \frac{d\Theta}{dt} + i[L,\Theta] \label{covt}\ee
Even when the variation is a symmetry, the holonomy $U$ is not invariant. Rather, it changes by
\be \delta U = iU(t_i,t_f)\Theta(t_f) - i \Theta(t_i)U(t_i,t_f)\ee
For cyclic paths, such that $x^i(t_i)=x^i(t_f)$, this means that $\delta U = [U,\Theta]$ which is the requirement that the holonomy of the vector space $V$ remains invariant up to a relabeling of the basis vectors of $V$. (Alternatively, up to a gauge transformation). In the remainder of this paper, we determine the constraints on Lagrangians $L$ which transform as a total covariant derivative \eqn{covt} under ${\cal N}=(2,2)$ supersymmetry.

\newpage
\subsection*{Chiral Multiplets and tt* Equations}

We first study connections $A$ which are functions of chiral multiplet parameters. The chiral multiplet consists of a complex scalar $\phi$, two complex Grassmann variables $\psi_+$ and $\psi_-$, and a complex auxiliary scalar $F$. The supersymmetry transformations are\footnote{Our spinor conventions are those of \cite{phases}, reduced to $d=0+1$ dimensions. To orient the reader with spinor contractions, it may help to recall that, in 4d, one can form a scalar $\psi\lambda$ and a 4-vector $\bar{\psi}\sigma^\mu\lambda$ from two Weyl spinors $\psi$ and $\lambda$. Upon dimensional reduction to $d=0+1$ dimensions, these descend to two scalars $\psi\lambda$ and $\bar{\psi}\lambda\equiv \bar{\psi}\sigma^0\lambda$, and a triplet of scalars $\bar{\psi}\vec{\sigma}\lambda$.},
\be
\delta \phi &=& \psi\epsilon\nn\\
\delta\psi_\pm &=& F\epsilon_\pm -i\dot{\phi}\bar{\epsilon}^\pm \label{chiral}\\
\delta F &=& -i\bar{\epsilon}\dot{\psi}\nn\ee
%
As a warm-up, we first construct a supersymmetric scalar-valued Lagrangian which starts with a connection term linear in time derivatives. We assign engineering dimensions consistent with the supersymmetry transformations: $[\phi]=0$, $[\lambda]=1/2$, $[F]=[d/dt]=1$ and $[\epsilon]=-1/2$. Then most general Lagrangian, at leading order in the derivative expansion, is given by,
\be L = A\dot{\phi} + A^\dagger \dot{\phi}^\dagger + GF + G^\dagger F  - \ft12B\psi\psi - \ft12 B^\dagger \bar{\psi}\bar{\psi}  + C\bar{\psi}\psi + \vec{C}\cdot\bar{\psi}\vec{\sigma}\psi \label{lag1}\ee
Here $A$, $G$ and $B$ are complex function of $\phi$ and $\bar{\phi}$, while $C$ is a real function and $\vec{C}$ is a triplet of real functions. $\vec{\sigma}$ are the Pauli matrices.

\para
We require that this Lagrangian is invariant under the supersymmetry transformations \eqn{chiral}. A direct computation gives constraints on the functions appearing in $L$: they must obey
\be \frac{\partial G}{\partial \phi^\dagger} = 0\ \ \ ,\ \ \ B = \frac{\partial G}{\partial \phi}\ \ \ ,\ \ \ \frac{\partial A}{\partial \phi^\dagger} -\frac{\partial A^\dagger}{\partial \phi} = 0\ \ \ ,\ \ \ C = \vec{C}=0\ee
With these restrictions, the Lagrangian is supersymmetric, transforming by a total derivative, 
$\delta L = \dot{\Theta}$, where
\be \Theta = A\psi\epsilon - iG\bar{\epsilon}\psi + {\rm h.c.}
\label{th1}\ee

\subsubsection*{\rm \underline{Matrix Valued Lagrangians}}

We now repeat the calculation, but this time with the Lagrangian \eqn{lag1} given by an $N\times N$ matrix. The functions $A$, $G$, $B$, $C$ and $\vec{C}$ are correspondingly promoted to 
$N\times N$ matrices.  Once more applying the supersymmetry transformations 
\eqn{chiral}, we insist that the Lagrangian transforms as a total covariant 
derivative \eqn{covt}, with $\Theta$ given by \eqn{th1}. We find that this 
imposes the constraints  $\vec{C}=0$ and,
\be {\cal D}^\dagger G &\equiv& \frac{\partial G}{\partial \phi^\dagger} + i[A^\dagger, G]=0 \nn\\
B &=& {\cal D}G \equiv \frac{\partial G}{\partial \phi}+i[A,\phi] \label{hitchin}\\ 
C &=& [G^\dagger,G] = [{\cal D},{\cal D}^\dagger]\nn\ee
These are the Hitchin equations \cite{Hitchin:1986vp} for the complex connection $A$ and 
complex matrix $G$. They arise as the double-dimensional reduction of the self-dual 
Yang-Mills equations. In the presence context, they can be thought of as a special case of the 
tt* equations. To derive the most general form of the tt* equations, we look at Lagrangians 
depending on several chiral multiplets.

\subsubsection*{\rm \underline{Multiple Chiral Multiplets}}

Consider multiple chiral multiplets, $(\phi^p,F^p,\psi^p_{\pm})$. The most general action matrix-valued Lagrangian that we can write down is,
\be L=(A_p\dot{\phi}^p +G_pF^p -B_{pq}\psi^p\psi^q +{\rm h.c.}) + C_{pq}\bar{\psi}^p\psi^q + \vec{C}_{pq}\cdot\bar{\psi}^p\vec{\sigma}\psi^q\ee
It is straightforward to vary this Lagrangian by the transformations \eqn{chiral}. 
Supersymmetry is assured if $\vec{C}_{pq}=0$ and
\be 
{\cal D}_p^\dagger G_q &\equiv& \frac{\partial G_q}{\partial \phi^{p\,\dagger}} + 
i[A_p^\dagger, G_q]=0  \nn\\
 {\cal D}_p G_q &=& D_q G_p =B_{pq} + B_{qp} \\
\left[G_p,G_q\right] &=& [{\cal D}_p,{\cal D}_q] = 0 \nn\\
C_{pq} &=& [G_p^\dagger,G_q] = [{\cal D}_q,{\cal D}_p^\dagger] 
\nn\ee
Here the covariant derivative is defined by ${\cal D}_p=\partial/\partial \phi_p + i[A_p,\cdot]$. 
These are the general form of the tt* equations of Cecotti and Vafa \cite{C1,tt*}. 
The original derivation of these equations came from studying the Berry connection 
in quantum mechanical systems; we will review this application shortly. The 
derivation presented here, invoking the invariance of a classical matrix Lagrangian, 
appears to be somewhat simpler.

\newpage
\subsection*{Vector Multiplets and Bogomolnyi Equations}

We now repeat the story for holonomies which depend on vector multiplet parameters. In $d=0+1$ dimensions, the vector multiplet consists of a single gauge field $a_0$, three real scalars $m^i$, two complex Grassmann variables $\lambda_\pm$, and a real auxiliary field $D$. (The scalars $m^i$ can be thought of as arising from the dimensional reduction of a vector field in $d=3+1$). The supersymmetry transformations are given by,
\be \delta a_0&=&i\bar{\lambda}\epsilon - i\bar{\epsilon}\lambda \nn\\
\delta \vec{m} &=& i\bar{\lambda}\vec{\sigma}\epsilon-i\bar{\epsilon}\vec{\sigma}\lambda \label{vector}\\
\delta \lambda &=& \dot{\vec{m}}\cdot\vec{\sigma}\epsilon+iD\epsilon\nn\\
\delta D &= &-\dot{\bar{\lambda}}\epsilon-\bar{\epsilon}\dot{\lambda}\nn\ee
We again start by considering the restrictions of supersymmetry on a scalar Lagrangian that starts with a connection term linear in time derivatives. In fact, this problem was already solved by Denef in \cite{denef}. The most general form of the Lagrangian is given by,
\be L = \vec{A}\cdot\dot{\vec{m}} - HD +B\lambda\lambda + B^\dagger \bar{\lambda}\bar{\lambda} + C\bar{\lambda}\lambda + \vec{C}\cdot\bar{\lambda}\vec{\sigma}\lambda\label{lag2}\ee
where $\vec{A}$, $H$, $C$ and $\vec{C}$ are real functions of $\vec{m}$, while $B$ is a complex function. A direct computation \cite{denef} shows that the transformations \eqn{vector} are a symmetry of this Lagrangian providing $B=C=0$ and
\be C_i=\frac{\partial H}{\partial m^i} = \epsilon_{ijk}\frac{\partial A_k}{\partial m^j}
\ee
With these restrictions, the Lagrangian transforms by a total derivative, $\delta L = \dot{\Theta}$, where
\be \Theta = H\bar{\lambda}\epsilon + i\vec{A}\cdot\bar{\lambda}\vec{\sigma}\epsilon +{\rm h.c.}\label{th2}\ee
\subsubsection*{\rm \underline{Matrix Valued Lagrangians}}

We now repeat this calculation for the vector multiplet Lagrangian \eqn{lag2}, with the functions $\vec{A}$, $H$, $C$, $\vec{C}$, and $B$ all promoted to $N\times N$ matrices. The calculation is once again straightforward. Applying the supersymmetry transformations \eqn{vector}, the Lagrangian transforms as a total covariant derivative \eqn{covt}, with $\Theta$ given by \eqn{th2}, providing that $B=C=0$ and,
\be C_i = {\cal D}_iH \equiv \frac{\partial H}{\partial m^i}+i[A_i,H] =  \ft12 \epsilon_{ijk}F_{jk}\label{bog}\ee
where the non-Abelian field strength is given by $F_{ij} = \partial_iA_j-\partial_jA_i+i[A_i,A_j]$. These are the Bogomolnyi monopole equations \cite{bog}. They arise as the dimensional reduction of the self-dual Yang-Mills equations. It is noteworthy that, for both chiral and vector multiplets, the constraints on the connections are related to the self-dual instanton equations.

\subsubsection*{\rm \underline{Multiple Vector Multiplets}}

Finally, we consider matrix-valued Lagrangians consisting of multiple vector multiplets
$(a_i^p,\vec{m}^p,D^p,\lambda_{\pm}^p)$. The most general action takes the form,
\be L=\vec{A}_p\dot{\vec{m}}^p - H_pD^p + (B_{pq}\lambda^p\lambda^q +{\rm h.c.}) +
C_{pq}\bar{\lambda}^p\lambda^q + \vec{C}_{pq}\cdot\bar{\lambda}^p\vec{\sigma}\lambda^q\ee
This time, supersymmetry requires $B=0$ and,
\be C_{pq} &=& [H_p,H_q]\nn\\
(C_i)_{pq} &=& \frac{\partial H_q}{\partial m^i_p} + i[A^i_p,H_q] =
\frac{\partial H_p}{\partial m^i_q} + i[A^i_q,H_p] \nn\\
\epsilon_{ijk}(C_k)_{pq} \!&-& i[H_p,H_q]\delta_{ij} =
\left(\frac{\partial A_p^j}{\partial m_q^i}-\frac{\partial A^i_q}{\partial m^j_p}
+i[A^i_q,A^j_p]\right)
\label{multiculti} \ee
These equations are to vector multiplets what the tt* equations are to chiral multiplets.

%
%

\subsection*{An Application: Berry Phase}

To end this paper, we describe an application of the above results to the computation of the Berry phase in strongly coupled quantum mechanical systems. Suppose that this quantum mechanics has $N$ degenerate ground states $|\,a\rangle$, $a=1,\ldots, N$, and let $x^i$ denote the parameters of the system. Then, as we adiabatically vary the parameters, the ground states will undergo a non-Abelian Berry holonomy \cite{berry,wz} given by \eqn{done}, where the $u(N)$ valued connection is
\be (A_i)_{ab} = i\langle b |\frac{\partial}{\partial
x^i} \,|a\rangle \label{conn}\ .\ee
Typically, the only way to compute the Berry connection is to first construct the ground states, and then use the direct definition \eqn{conn}. However, in supersymmetric quantum mechanics, once can bypass this step. In many examples, this allows the Berry connection to be computed exactly, even when the ground states cannot be. 
The key point is that the parameters in supersymmetric theories themselves sit in supermultiplets. One can integrate out the all dynamical fields to get an effective $u(N)$-valued Lagrangian for the parameters of the form \eqn{lag}. This Lagrangian must itself be invariant under supersymmetry. We now give some examples of this procedure.

\subsubsection*{\underline{\rm Chiral Multiplets}}

Consider a Wess-Zumino model in $d=0+1$ dimensions. The superpotential depends on the dynamical chiral multiplets, which we collectively call $Y$, and the complex parameters $\phi$: ${\cal W} = {\cal W}(Y;\phi)$.

\para
Supersymmetric ground states are defined by $\partial{\cal W}/\partial Y=0$. We are interested in how the space of ground states varies as one changes the parameters $\phi$. This is precisely the information captured by the Berry connection \eqn{conn}. The requirement that the effective action for the parameters $\phi$ is supersymmetric, ensures that the Berry connection must satisfy the tt* equations \eqn{hitchin}. We need only understand the meaning of the complex matrix $G$ in the original quantum mechanics. Expanding out the superpotential, the  auxiliary field $F$ --- which is the superpartner of the parameter $\phi$ ---  appears in the quantum mechanical Lagrangian as,
\be \int d^2\theta\ {\cal W}(Y;\phi) = \frac{\partial {\cal W}}{\partial \phi}F+\ldots\ee
$F$ can be viewed as a source in the original quantum mechanics, such that differentiating with respect to $F$ computes the expectation value of $\partial {\cal W}/\partial \phi$. Comparing with the effective action \eqn{lag1} for the parameters, we see that
\be\label{eq:ringcoeffs} (G)_{ab} = \langle b | \frac{\partial {\cal W}}{\partial \phi} |a\rangle\ee
The tt* equations then relate the curvature of the Berry connection to this matrix element.

\para
The result that the Berry phase in Wess-Zumino models obeys the tt* equations is certainly not new; it was one of the main results of the original tt* papers \cite{C1,tt*}, where the matrix elements (\ref{eq:ringcoeffs}) were interpreted as the coefficients of the chiral ring. Nonetheless, the derivation given here, in terms of a non-Abelian effective action for the parameters, appears to be novel.

\subsubsection*{\underline{\rm Vector Multiplets}}

There is a similar story for parameters which live in ${\cal N}=(2,2)$ vector multiplets. Consider a $d=0+1$ supersymmetric sigma-model with dynamical fields $Y$ and $\zeta$,
\be L_{\sigma-{\rm model}}= \ft12 g_{mn}(Y)\dot{Y}^m\dot{Y}^n + g_{mn}\bar{\zeta}^mD_t\zeta^n + R_{mnpq}\bar{\zeta}^m\zeta^n\bar{\zeta}^p\zeta^q
\ee
where $D_t\zeta^n = \dot{\zeta}^n+\Gamma^n_{pq}\dot{Y}^p\zeta^q$. As is well known, ${\cal N}=(2,2)$ supersymmetry requires the the metric $g$ is K\"ahler. If the metric admits a holomorphic Killing vector $k_m$, then one may add a potential over the target space which depends on three parameters $\vec{m}$,
\be L_{\rm potential} = |\vec{m}|^2\, k^2 + g_{mn}\bar{\zeta}^m(\vec{m}\cdot\vec{\sigma})\zeta^n\ee
The parameters $\vec{m}$ live in a background vector multiplet.

\para
The Witten index for this system guarantees the existence of at least $Tr(-1)^F=N$ ground states, where $N$ is the Euler character of the target space. (This statement is true only for compact target spaces). 
We want to know the Berry connection \eqn{conn} for these grounds states as the parameters $\vec{m}$ are varied.  We may again integrate out the dynamical degrees of freedom $Y$ and $\zeta$, to leave ourselves with an effective action of the form \eqn{lag2}. The results above tells us that the Berry connection must satisfy the Bogomolnyi monopole equation \eqn{bog}.
%
%

\para
It remains to determine the matrix $H$ that appears in the Bogomolnyi equation in terms of the original dynamical variables $Y$. To do this, we must understand how $D$, the auxiliary superpartner of $\vec{m}$, couples to the system. This can be read off from \cite{bw}. There is a term in the sigma-model Lagrangian proportional to $\mu(Y) D$,
where $\mu(Y)$ is the moment map associated to the Killing vector $k$. This is a function over the target space which satisfies $d\mu =  \imath_{k}\omega$, where $\omega$ is the K\"ahler form. We therefore find that
\be H_{ab} = \langle b | \mu(Y) |a \rangle \label{above}\ee
In \cite{us1,us2}, we studied the ${\bf CP}^1$ sigma-model and, by explicit computation, showed that the Berry connection was given by the single $SU(2)$ BPS monopole satisfying \eqn{bog}. We conjectured that the Berry phase for the ${\bf CP}^{N-1}$ sigma-model was the $(1,1,\ldots,1)$ BPS monopole in $SU(N)$ gauge theory. The results of this paper prove this conjecture.

\para
Recently, it was proposed that Berry phases could be used to manipulate the microstates of supersymmetric black holes \cite{bh1,bh2}. The idea was to consider how the microstates of the black hole change as one varies expectation values for the asymptotic scalars. These appear as parameters in the black hole quantum mechanics, and the Berry connection can be shown to satisfy a modification of the tt* equations \cite{bh1,bh2}. In fact, the non-Abelian monopole connections \eqn{multiculti} also appear to be relevant in this context. The scalars in the vector multiplet parameterize the separation of multi-centered black holes in four-dimensions \cite{denef}. The non-Abelian monopole connections that we find in this paper describe the holonomy of microstates as black holes orbit in the Born-Oppenheimer approximation. It would be interesting to explore this connection further.

\newpage

\acknowledgments{We are grateful to Kentaro Hori for a vital conversation. This work is supported in part by funds provided by the U.S. Department of Energy (D.O.E.) under cooperative research agreement DEFG02-05ER41360. J.S. is supported by NSF grant PHY-0600465 and thanks the members of CTP at MIT, and especially Dan Freedman, for hospitality. D.T. is supported by the Royal Society.}


\end{document}